   \documentclass[12pt,preprint]{aastex}
\shorttitle{Modified equipartition calculation for supernova
remnants} \shortauthors{Arbutina et al.}

\begin{document}

\title{Modified equipartition calculation for supernova
remnants}

\author{B. Arbutina$^1$, D. Uro\v{s}evi\'{c}$^{1,2}$, M. M. Andjeli\'{c}$^1$, M. Z. Pavlovi\'{c}$^1$ and B. Vukoti\'{c}$^3$}
\affil{$^1$Department of Astronomy, Faculty of Mathematics, University
of Belgrade, Studentski trg 16, 11000 Belgrade, Serbia}
\email{arbo@math.rs}
\affil{$^2$Isaac Newton Institute of Chile, Yugoslavia Branch}
\affil{$^3$Astronomical Observatory, Volgina 7, 11060 Belgrade 38, Serbia}

%\altaffiltext{1}{Isaac Newton Institute of Chile, Yugoslavia Branch}

\begin{abstract}
Determination of the magnetic field strength in the interstellar medium is one of the most complex
tasks of contemporary astrophysics. We can only estimate the order of magnitude of the magnetic field strength by using a few very
limited methods. Besides Zeeman effect and Faraday rotation,
the equipartition or the minimum-energy calculation is a widespread
method for estimating magnetic field strength and energy contained in the
magnetic field and cosmic ray particles by using only the radio
synchrotron emission. Despite of its approximate character, it
remains a useful tool, especially when there is no other data about the magnetic field in a
source. In this paper we give a modified calculation
which we think is more appropriate for estimating magnetic field strengths
and energetics in supernova remnants (SNRs). Finally, we present calculated estimates of the magnetic field strengths for all Galactic SNRs for which the necessary observational data are available. The web application for calculation of the magnetic field strength of SNRs is available at {\it http://poincare.matf.bg.ac.rs/\~{}arbo/eqp/}.
\end{abstract}


\section{Introduction}

The basic constituents of the interstellar medium (ISM) are: normal (thermalized) particles, cosmic rays (CRs), radiation and magnetic field. Each of these four forms of ISM contains similar energy density of about 1 eV/cm$^3$. If we compare quantity of information available for each of them, we can immediately conclude that the magnetic field is absolutely the most intrigued and hidden form of ISM. Recent simulations of SNR shocks commonly include magnetic field because it plays an important part in various related phenomena (particle acceleration, radiation, shock compression and formation, etc). The magnetic field strength and its direction can only be approximately estimated by using a few, in their applicabilities, very limited methods (for recent review of magnetic fields in supernova remnants see Reynolds et al. 2011). One of them is Zeeman effect - it is appropriate method for generally stronger fields - it can be used for determination of strong ISM magnetic fields in high density HI or molecular clouds rich with OH and CN. The global magnetic field of the Galaxy, a few $\mu$G, is too small to be measured in this way. The second method for determination of the component of ISM magnetic field parallel to the line of sight is so-called Faraday rotation or rotation measure method.
Rotation measure (RM) is calculated directly from the radio astronomical
polarization observations at multiple frequencies. This quantity depends on
the plasma density and the strength of the field component along the line of
sight. Under necessary simplistic
assumptions RM can yield an order of
magnitude estimate of the magnetic field strength between the source and
observer. If several distinct rotating regions located
along the line of sight generate a spectrum of various RM components, multi-channel spectro-polarimetric radio data are needed that can be
Fourier-transformed into
Faraday space, called RM synthesis (see Heald 2009, Beck 2011 and references therein).
 If we would like to estimate the magnetic field strength directly connected to a source embedded in the relatively low density region, the only way is by using the so-called equipartition calculation.

The equipartition or the minimum-energy calculation is a
widespread method for estimating magnetic field strength and
energy contained in the magnetic field and cosmic ray particles by
using only the radio synchrotron emission of a source. Despite of
its approximate character, it remains a useful tool in situations
when no other data about the source are available. Details of
equipartition and revised equipartition calculations for radio
sources in general are available in Pacholczyk (1970, hereafter
P70), Govoni \& Feretti (2004), and Beck \& Krause (2005, hereafter BK05), respectively. A
discussion on whether equipartition of energy is fulfilled in real
sources, and how reliable magnetic field estimates from
equipartition calculation are, can be found in Duric (1990).

In his famous book, Pacholczyk gave fundamental concepts of the
equipartition or the minimum-energy calculation. The first
ingredient of the equipartition calculation is expression for
total energy of relativistic particles, which can be obtained by
integration of power-law energy distribution of cosmic rays. Total
energy of relativistic particles was found by integration over all
frequencies in the radio domain. Pacholczyk assumed homogenous
magnetic field for calculation of energy contained in the magnetic
field, and coefficient $k$ which represents ratio between energies
of relativistic protons and electrons. Finally, the last
ingredient in the P70 equipartition formula is the radio
luminosity of an object.

BK05 presented the revised equipartition calculation. The basic
improvement in comparison to the classical, P70 equipartition, is
integration of power-law energy distribution over energies instead
over frequencies. They integrated over two energy ranges with a
break at $E=mc^2$ where $m$ is the rest mass of the accelerated
particles, i.e. two power-law distributions with different slopes,
both dependent on energy spectral index $\gamma$. Instead of luminosity
used in the classical approach, BK05 used radio intensity -
their intention was to determine the magnetic field strength of
the small part of the very extended objects such as whole Galaxy
or an extragalactic system. The magnetic field small scale
structures of very extended objects are very far from being
homogenous. The model of magnetic field distribution used in the
revised equipartition formula is accommodated for the previously
described objects. Finally, BK05 used coefficient $\mathbf{K_0}$ which
represents ratio of the number densities of cosmic ray protons to
electrons, instead of ratio between energies of protons and
electrons used in the classical equipartition.

In this paper, we use the energy ratio, as in the classical
calculation, but it includes all heavier particles which can be
found in cosmic rays. Also, we use the radio flux density instead
the radio luminosity as in P70 equipartition, or the specific
intensity from revised calculation of BK05. Since our intention is
to derive equipartition formulae for the determination of the
magnetic fields and the minimal energies    in supernova remnants
(SNRs), we use model of the magnetic field distribution defined in
Longair (1994). Finally, since the distribution of CRs is a
power-law in momentum (which can be transformed to the same
power-law in energy, for energies high enough), we have chosen to
integrate over momentum and not over energies as BK05 did, so
there is no need for introduction of the break in the differential
energy
 spectrum.

We emphasize that the final formulae in the P70 equipartition do
not depend on the energy spectral index (or radio spectral index ,
$\alpha=(\gamma-1)/2$), while in the BK05 and our equipartition
these formulae depend on the energy spectral index (see equations (12) and
(13)).

In the next section, by relying on Bell's theory of diffusive
shock acceleration - DSA, (Bell 1978a,b), and his assumption
concerning injection of particles into the acceleration process,
we will first derive a modified equipartition i.e. minimum-energy
calculation (Arbutina et al. 2011) applicable to 'mature' SNRs
($v_s \ll 6000 - 7000 $ km/s) with radio spectral index $0.5<\alpha <1$
(energy spectral index $2< \gamma <3$). Then we will incorporate the
dependence $\epsilon = \epsilon (E_{\mathrm{inj}})$ which will
make formula applicable to the younger i.e. all SNRs.

\section{Analysis and Results}

\subsection{A simple approach}

Following Bell (1978b) we will assume that a certain number of
particles have been injected into the acceleration process all
with the same injection energy $E_{inj} \approx 4 \frac{1}{2} m_p
v_s^2$.\footnote{We assume fully ionized, globally electro-neutral
plasma.} If we assume that shock velocity is low enough so that
$E_{\mathrm{inj}} \ll m_e c^2$ (and $p^e_{\mathrm{inj}} \ll m_e
c$), for energy density of a cosmic ray species (e.g. electrons, protons, $\alpha$-particles, heavier ions), assuming
power-law momentum distribution, we have
\begin{eqnarray}
\epsilon  &=& \int _{p_{\mathrm{inj}}} ^{p_\infty} 4\pi k
p^{-\gamma} (\sqrt{p^2c^2+m^2c^4}-mc^2)\mathrm{d}p \nonumber
\\
&\approx & \int _0 ^{\infty} 4\pi k p^{-\gamma}
(\sqrt{p^2c^2+m^2c^4}-mc^2)\mathrm{d}p \nonumber
\\
&= & 4\pi k c (mc)^{2-\gamma} \int _0 ^{\infty} x^{-\gamma}
(\sqrt{x^2 + 1}-1)\mathrm{d}x, \ \ \ x=\frac{p}{mc} \nonumber
\\
 &= & K
(mc^2)^{2-\gamma}\frac{\Gamma (\frac{3-\gamma}{2})\Gamma
(\frac{\gamma -2}{2}) }{2\sqrt{\pi}(\gamma -1)},\ \ \
 K= 4\pi kc^{\gamma -1},\ \ \ 2< \gamma <3.
\end{eqnarray}
where $k$ is the constant in the distribution function $f(p) = k p^{-(\gamma +2)}$.
Function under the integral in equation (1) is approximately a power-law
with a spectral index of $2-\gamma$ for thermal (non-relativistic) particles and a
power-law with a spectral index of $1-\gamma$ for highly relativistic particles.
In this paper the sharp break in BK05 is replaced by a smooth one.

Total cosmic ray energy density is then
\begin{eqnarray}
\epsilon _{\mathrm{CR}} &=& \frac{\Gamma
(\frac{3-\gamma}{2})\Gamma (\frac{\gamma -2}{2})
}{2\sqrt{\pi}(\gamma -1)} \Bigg(K_e (m_ec^2)^{2-\gamma}+\sum _i
K_i
(m_i c^2)^{2-\gamma}\Bigg) \nonumber \\
& =&  \frac{\Gamma (\frac{3-\gamma}{2})\Gamma (\frac{\gamma
-2}{2}) }{2\sqrt{\pi}(\gamma -1)} \Bigg(K_e (m_ec^2)^{2-\gamma}+
K_p (m_p c^2)^{2-\gamma} \sum _i
\frac{n_i}{n_p}  \Big(\frac{m_i}{m_p}\Big)^{(3-\gamma)/2}  \Bigg) \nonumber \\
& =&  \frac{\Gamma (\frac{3-\gamma}{2})\Gamma (\frac{\gamma
-2}{2}) }{2\sqrt{\pi}(\gamma -1)} K_e (m_e c^2)^{2-\gamma} \Bigg(1
+ \frac{n}{n_e} \Big(\frac{m_p}{m_e}\Big)^{(3-\gamma)/2} \sum _i
\frac{n_i}{n}  \Big(\frac{m_i}{m_p}\Big)^{(3-\gamma)/2}  \Bigg) \nonumber \\
& =&  K_e (m_e c^2)^{2-\gamma}\frac{\Gamma
(\frac{3-\gamma}{2})\Gamma (\frac{\gamma -2}{2})
}{2\sqrt{\pi}(\gamma -1)} (1+\kappa),
\end{eqnarray}
where
\begin{equation}
\kappa = \Big(\frac{m_p}{m_e}\Big)^{(3-\gamma)/2} \frac{\sum _i
A_i ^{(3-\gamma)/2}\nu _i }{\sum _i Z_i \nu _i},
\end{equation}

\noindent $\kappa$ represents the energy ratio between ions and electrons,
$n_e = \sum _i Z_i n_i$, $\nu _i =n_i /n$ are ion abundances,
$A_i$ and $Z_i$ are mass and charge numbers of elements and we
assumed that at high energies $K_p/K_e \approx (n_p/n_e)\cdot$
$({m_p}/{m_e})^{(\gamma -1)/2}$ (see equation (26)), where $K_p$ and $K_e$ are the constants in the power-law energy distributions for protons and electrons, respectively. Note that we have neglected
energy losses.

Emission coefficient for synchrotron radiation is, on the other
hand,
\begin{equation}
\varepsilon _\nu = c_5 K_e (B \sin \Theta )^{(\gamma +1)/2}
\Big(\frac{\nu}{2c_1}\Big)^{(1-\gamma )/2},
\end{equation}
where $c_1, c_3$ and $c_5 = c_3 \Gamma (\frac{3\gamma -1}{12})
\Gamma (\frac{3\gamma +19}{12})/(\gamma +1)$ are defined in
P70.\footnote{Namely, $c_1= 6.264\cdot 10^{18}$ and
$c_3= 1.866\cdot 10^{-23}$ in cgs units.} We will use the flux
density which is defined as
\begin{equation}
S _\nu = \frac{L_\nu}{4\pi d^2} = \frac{\mathcal{E}_\nu V}{4\pi d^2} = \frac{\frac{4\pi}{3}R^3 f
\mathcal{E}_\nu}{4\pi d^2} = \frac{4\pi}{3} \varepsilon _\nu f
\theta ^3 d,
\end{equation}
where $L_\nu$ is radio luminosity, $\mathcal{E}_\nu$ is volume emisivity, $V$ is the volume, $f$ is volume filling factor of radio emission, $R$ is the radius, $d$ is the distance and $\theta = R/d$ is angular radius.

If we assume isotropic distribution for the orientation of
pitch angles (Longair 1994) we can take for the average
$\langle{(\sin \Theta )^{(\gamma +1)/2}}\rangle $
\begin{equation}
 \frac{1}{2} \int
_0^\pi (\sin \Theta )^{(\gamma +3)/2} \mathrm{d}\Theta =
\frac{\sqrt{\pi}}{2}\frac{\Gamma (\frac{\gamma +5}{4})}{\Gamma
(\frac{\gamma +7}{4})}.
\end{equation}

For the total energy we have
\begin{equation}
E = \frac{4\pi}{3}R^3 f (\epsilon _{\mathrm{CR}} + \epsilon _B), \
\ \ \epsilon _B = \frac{1}{8\pi} B^2,
\end{equation}
\begin{equation}
E = \frac{4\pi}{3}R^3 f \Bigg( K_e (m_e
c^2)^{2-\gamma}\frac{\Gamma (\frac{3-\gamma}{2})\Gamma
(\frac{\gamma -2}{2}) }{2\sqrt{\pi}(\gamma -1)} (1+\kappa) +
\frac{1}{8\pi} B^2\Bigg).
\end{equation}
Looking for the minimum energy with respect to $B$,
$\frac{\mathrm{d}E}{\mathrm{d}B} =0$ gives
\begin{equation}
\frac{\mathrm{d} K_e}{\mathrm{d}B} (m_e
c^2)^{2-\gamma}\frac{\Gamma (\frac{3-\gamma}{2})\Gamma
(\frac{\gamma -2}{2}) }{2\sqrt{\pi}(\gamma -1)} (1+\kappa) +
\frac{1}{4\pi} B =0,
\end{equation}
where (by using (4), (5) and (6))
\begin{equation}
\frac{\mathrm{d} K_e}{\mathrm{d}B} = - \frac{3}{4\pi}
\frac{S_\nu}{f \theta ^3 d }\frac{1}{c_5}
\Big(\frac{\nu}{2c_1}\Big)^{-(1-\gamma )/2} \frac{(\gamma +1)
\Gamma (\frac{\gamma +7}{4})}{\sqrt{\pi}\Gamma (\frac{\gamma
+5}{4})} B^{-(\gamma +3)/2},
\end{equation}
i.e. the magnetic field for the minimum energy is
\begin{eqnarray}
B = \Big(\frac{3}{2\pi}\frac{(\gamma +1)\Gamma
(\frac{3-\gamma}{2})\Gamma(\frac{\gamma -2}{2})\Gamma
(\frac{\gamma +7}{4})}{(\gamma -1)\Gamma (\frac{\gamma +5}{4})}
 \frac{S_\nu}{f d \theta ^3} \cdot
\nonumber \\
\cdot (m_e c^2)^{2-\gamma} \frac{(2c_1)^{(1-\gamma)/2}}{c_5}
(1+\kappa) \nu ^{(\gamma -1)/2}\Big)^{2/(\gamma +5)},
\end{eqnarray}
or
\begin{eqnarray}
 B\ \mathrm{[G]} &\approx & \Big(6.286\cdot 10^{(9\gamma -79)/2}
\frac{\gamma +1}{\gamma -1}\frac{\Gamma
(\frac{3-\gamma}{2})\Gamma(\frac{\gamma -2}{2})\Gamma
(\frac{\gamma +7}{4})}{\Gamma (\frac{\gamma +5}{4})} (m_e
c^2)^{2-\gamma}
 \cdot   \\
 &\cdot &  \frac{(2c_1)^{(1-\gamma)/2}}{c_5}  (1+\kappa)  \frac{S_\nu \mathrm{[Jy]}}{f\ d
 \mathrm{[kpc]}\
 \theta
[\mathrm{arcmin}]^3} \nu[\mathrm{GHz}] ^{(\gamma
-1)/2}\Big)^{2/(\gamma +5)}, \nonumber
\end{eqnarray}
where $m_e c^2 \approx 8.187\cdot 10^{-7}$ ergs. We also have
\begin{equation}
E_B = \frac{\gamma +1}{4} E_{\mathrm{CR}}, \ \ \ E_{\mathrm{min}}
= \frac{\gamma +5}{\gamma +1} E_B.
\end{equation}
This result is the same as in BK05.

\subsection{A more general formula for $\kappa$}

Let us start again with equation (1)

\begin{eqnarray}
\epsilon &\approx & \int _{p_{\mathrm{inj}}} ^{\infty} 4\pi k
p^{-\gamma} (\sqrt{p^2c^2+m^2c^4}-mc^2)\mathrm{d}p \nonumber
\\
&= & 4\pi k c (mc)^{2-\gamma} \int _{\frac{p_{\mathrm{inj}}}{mc}}
^{\infty} x^{-\gamma} (\sqrt{x^2 + 1}-1)\mathrm{d}x, \ \ \
x=\frac{p}{mc} \nonumber
\\
 &= & 4\pi k c (mc)^{2-\gamma} I\Big(\frac{p_{\mathrm{inj}}}{mc}\Big).
\end{eqnarray}
Integral $I(x)$ can be expressed through Gauss hypergeometric
function $_2F_1$ (for $\gamma >2$)
\begin{equation}
I(x) = \frac{\Gamma (\frac{3-\gamma}{2})\Gamma (\frac{\gamma
-2}{2}) }{2\sqrt{\pi}(\gamma -1)} - \frac{x^{1-\gamma}(1-\ _2F_1
(-\frac{1}{2}, \frac{1-\gamma}{2}, \frac{3-\gamma}{2};
-x^2))}{\gamma -1},
\end{equation}
but we will try to find more simple approximation. First notice
that
\begin{eqnarray}
& I(x) \approx \frac{\Gamma (\frac{3-\gamma}{2})\Gamma
(\frac{\gamma
-2}{2}) }{2\sqrt{\pi}(\gamma -1)} - \frac{x^{3-\gamma}}{2(3-\gamma)} + \frac{x^{5-\gamma}}{8(5-\gamma)}-\ldots, \ \ \ x\rightarrow 0, \\
& I(x) \approx \frac{x^{2-\gamma}}{\gamma -2} , \ \ \ x\rightarrow
\infty.
\end{eqnarray}
So we can try an approximation ($2< \gamma <3$)
\begin{equation}
I(x)_\mathrm{approx} = \frac{\frac{\Gamma
(\frac{3-\gamma}{2})\Gamma (\frac{\gamma -2}{2})
}{2\sqrt{\pi}(\gamma -1)} - \frac{x^{3-\gamma}}{2(3-\gamma)} +
F(\gamma)x^{5-\gamma}}{1+F(\gamma) (\gamma -2)x^3}
\end{equation}
which has correct limits when $x\rightarrow 0$ and $x\rightarrow
\infty$. We shall find $F(\gamma)$ from matching condition $I(1)
=I(1)_\mathrm{approx}$:
\begin{equation}
F(\gamma) = \frac{\frac{1}{2(3-\gamma)} - \frac{1-\ _2F_1
(-\frac{1}{2}, \frac{1-\gamma}{2}, \frac{3-\gamma}{2}; -1)}{\gamma
-1} }{1- (\gamma -2) (\frac{\Gamma (\frac{3-\gamma}{2})\Gamma
(\frac{\gamma -2}{2}) }{2\sqrt{\pi}(\gamma -1)}   - \frac{1-\
_2F_1 (-\frac{1}{2}, \frac{1-\gamma}{2}, \frac{3-\gamma}{2};
-1)}{\gamma -1} )}.
\end{equation}
Since the last expression also involves hypergeometric function we
found by trial and error an approximation
\begin{equation}
F(\gamma)_\mathrm{approx} = \frac{17}{1250}\frac{(2\gamma
+1)\gamma}{(\gamma -2)(5-\gamma)}
\end{equation}
From now on we will assume $I(x) =I(x)_\mathrm{approx}$ and
$F(\gamma) = F(\gamma)_\mathrm{approx}$ (relative error is less
than 3.5 \%).

Total cosmic rays energy density is then
\begin{eqnarray}
\epsilon _{\mathrm{CR}} &=&  \epsilon _e + \epsilon _\mathrm{ion}
= K_e (m_ec^2)^{2-\gamma}
I\Big(\frac{p^e_\mathrm{inj}}{m_ec}\Big)+\sum _i K_i (m_i
c^2)^{2-\gamma} I\Big(\frac{p^i_\mathrm{inj}}{m_i c}\Big) ,
\end{eqnarray}
where (because $\frac{p^i_\mathrm{inj}}{m_i c} \ll 1$)
\begin{eqnarray}
\epsilon _\mathrm{ion} &\approx& \sum _i K_i (m_i c^2)^{2-\gamma}
\Bigg( \frac{\Gamma (\frac{3-\gamma}{2})\Gamma (\frac{\gamma
-2}{2}) }{2\sqrt{\pi}(\gamma -1)} -\frac{1}{2(3-\gamma)} \Bigg( \frac{\sqrt{E_\mathrm{inj}^2+2m_i c^2E_\mathrm{inj}}}{m_ic^2} \Bigg)^{3-\gamma} \Bigg) \nonumber \\
& \approx & K_p (m_p c^2)^{2-\gamma} \sum _i
\frac{n_i}{n_p} \Big( \frac{p^i_\mathrm{inj}}{p^p_\mathrm{inj}}\Big)^{\gamma -1} \Big(\frac{m_i}{m_p}\Big)^{2-\gamma} \cdot \nonumber \\
&\cdot& \Bigg( \frac{\Gamma (\frac{3-\gamma}{2})\Gamma
(\frac{\gamma -2}{2}) }{2\sqrt{\pi}(\gamma -1)}
-\frac{1}{2(3-\gamma)} \Bigg( \frac{2E_\mathrm{inj}}{m_ic^2} \Bigg)^{(3-\gamma)/2} \Bigg) \nonumber \\
& \approx & K_p (m_p c^2)^{2-\gamma} \sum _i \Bigg[
\frac{n_i}{n_p} \Big(\frac{m_i}{m_p}\Big)^{(3-\gamma)/2}
\frac{\Gamma (\frac{3-\gamma}{2})\Gamma (\frac{\gamma -2}{2})
}{2\sqrt{\pi}(\gamma -1)}
-\frac{1}{2(3-\gamma)} \Bigg( \frac{2E_\mathrm{inj}}{m_pc^2} \Bigg)^{(3-\gamma)/2} \frac{n_i}{n_p} \Bigg] \nonumber \\
& \approx & K_p (m_p c^2)^{2-\gamma} \frac{n}{n_p} \Bigg[
\frac{\Gamma (\frac{3-\gamma}{2})\Gamma (\frac{\gamma -2}{2})
}{2\sqrt{\pi}(\gamma -1)} \sum _i A_i ^{(3-\gamma)/2}\nu _i
-\frac{1}{2(3-\gamma)} \Bigg( \frac{2E_\mathrm{inj}}{m_pc^2}
\Bigg)^{(3-\gamma)/2} \Bigg] .
\end{eqnarray}

Finally
\begin{eqnarray}
\epsilon _{\mathrm{CR}} & =& K_e (m_e c^2)^{2-\gamma} \Bigg[
I\Bigg( \frac{\sqrt{E_\mathrm{inj}^2+2m_e c^2E_\mathrm{inj}}}{m_e
c^2} \Bigg)  + \frac{1}{\sum _i Z_i \nu _i} \Big(
\frac{m_p}{m_e}\Big)^{2-\gamma} \Big( \frac{
2m_pc^2E_\mathrm{inj}}{E_\mathrm{inj}^2 + 2m_ec^2E_\mathrm{inj}}
\Big)^{(\gamma -1)/2} \cdot \nonumber \\
&\cdot &
 \Bigg( \frac{\Gamma
(\frac{3-\gamma}{2})\Gamma (\frac{\gamma -2}{2})
}{2\sqrt{\pi}(\gamma -1)} \sum _i A_i ^{(3-\gamma)/2}\nu _i
-\frac{1}{2(3-\gamma)} \Bigg( \frac{2E_\mathrm{inj}}{m_pc^2}
\Bigg)^{(3-\gamma)/2}  \Bigg) \Bigg]
\nonumber \\
&=& K_e (m_e c^2)^{2-\gamma}\frac{\Gamma
(\frac{3-\gamma}{2})\Gamma (\frac{\gamma -2}{2})
}{2\sqrt{\pi}(\gamma -1)} (1+\kappa),
\end{eqnarray}
where
\begin{eqnarray}
\kappa & =&   I\Bigg( \frac{\sqrt{E_\mathrm{inj}^2+2m_e c^2
E_\mathrm{inj}}}{m_e c^2} \Bigg)   \Bigg( \frac{\Gamma
(\frac{3-\gamma}{2})\Gamma (\frac{\gamma -2}{2})
}{2\sqrt{\pi}(\gamma -1)} \Bigg)^{-1} + \frac{1}{\sum _i Z_i \nu
_i} \Big( \frac{m_p}{m_e}\Big)^{2-\gamma} \Big( \frac{
2m_pc^2E_\mathrm{inj}}{E_\mathrm{inj}^2 +
2m_ec^2E_\mathrm{inj}} \Big)^{(\gamma -1)/2} \cdot \nonumber \\
&\cdot &  \Bigg(
 \sum _i A_i ^{(3-\gamma)/2}\nu _i
-\frac{1}{2(3-\gamma)} \Bigg( \frac{2E_\mathrm{inj}}{m_pc^2}
\Bigg)^{(3-\gamma)/2}   \Bigg( \frac{\Gamma
(\frac{3-\gamma}{2})\Gamma (\frac{\gamma -2}{2})
}{2\sqrt{\pi}(\gamma -1)} \Bigg)^{-1} \Bigg)  -1.
\end{eqnarray}

In the above derivation we used the fact that (Bell 1978b)
\begin{equation}
K_i/K_p = \frac{n_i}{n_p}
\Big(\frac{p^i_\mathrm{inj}}{p^p_\mathrm{inj}}\Big)^{\gamma -1}
\approx (n_i/n_p)\cdot ({m_i}/{m_p})^{(\gamma -1)/2}
\end{equation}
and
\begin{equation}
K_p/K_e = (n_p/n_e) \Big( \frac{E_\mathrm{inj}^2 +
2m_pc^2E_\mathrm{inj}}{E_\mathrm{inj}^2 + 2m_ec^2E_\mathrm{inj}}
\Big)^{(\gamma -1)/2} \approx (n_p/n_e)\cdot \Big(\frac{2m_p
c^2E_\mathrm{inj}}{E_\mathrm{inj}^2 +
2m_ec^2E_\mathrm{inj}}\Big)^{(\gamma -1)/2}.
\end{equation}
Equation (24) has the correct limit (3) when $E_{\mathrm{inj}} \ll
m_e c^2 \ll  m_p c^2$. From Figure 1 it can be seen that for low
$E_{\mathrm{inj}}$ cosmic rays energy density is almost constant
(independent of $E_{\mathrm{inj}}$) and usage of equation (3) is
justified. When shock velocity can be estimated one should
calculate injection energy $E_{\rm inj} \approx 4 \frac{1}{2} m_p
v_s^2$ and use equation (24). Formulae (12) and (13) for magnetic field
and minimum energy remain the same.\footnote{Note that $\kappa$ is no longer ions to electrons energy ratio but a suitable parameter
introduced to make new formulae same as the old ones.} In Figure 2 we give proton to
electron energy density ratio as a function of injection energy in
our approximation compared to the same data from Bell (1978b).
Agreement is quite good despite the approximative character of our
formulae.

We have implemented our modified equipartition calculation by
developing a PHP code.\footnote{The calculator is available at
{\it http://poincare.matf.bg.ac.rs/\~{}arbo/eqp/}.} The code uses
some 'typical' starting values for radio spectral index, frequency, flux
density, distance, angular radius, filling factor, shock velocity
and abundances, which all can be changed or left as such. For
example, if shell thickness relative to SNR radius $\delta$ can be
measured the volume filling factor is $f = 1 - (1-\delta)^3$.
Otherwise a typical value $f=0.25$ can be used (shell thickness of
about 10 \%). If shock velocity is unknown one should leave
0 (and simpler equipartition calculation will be performed by
using equation (3)). Simple ISM abundances are
assumed for start (H:He ratio 10:1). In the implementation of our
calculation we used an approximation for the Gamma function (Nemes
2010):
\begin{equation}
\Gamma (z) = \sqrt{\frac{2\pi}{z}}\Bigg( \frac{1}{e}\Bigg(z+
\frac{1}{12z-\frac{1}{10z}} \Bigg) \Bigg)^z.
\end{equation}

\section{Discussion}

From the mathematical point of view, the equipartition calculation
is the problem of solving a system of two independent equations
(the synchrotron emissivity equation (4) and equation for the
total energy in a source (7)) for the three unknown variables (the
total energy $E$, energy contained in the cosmic rays $E_{\rm CR}$
(or $K_{\rm e}$), and energy contained in the magnetic field $E_B$
(or $B$)). This problem is, of course, impossible to solve without
additional assumption. The primary assumption is to seek for the
minimum of the total energy of the synchrotron source.
Differentiation of equation (8) make that the total energy
disappears as unknown variable and two starting equations ((4) and
(9)) can now give us solutions for both remaining unknown
variables ($K_{\rm e}$ and $B$). As the result of differentiation
of equation (8), the exact equipartition between energies
contained in the magnetic field and cosmic rays is only
approximately fulfilled (equation (13)). The alternative
assumption, commonly adopted, is the equipartition between  energy
contained in the cosmic rays and in the magnetic field ($\epsilon_{\rm
CR}=\epsilon_B$). By assuming this we directly link  $K_{\rm e}$ and $B$
(see equations (2) and (7)). It is often the case in the literature that these two calculations are commonly referred
as either equipartition or the minimum energy calculation.
Here, we would like to emphasize that
strict equipartition does not have to be assumed for doing the
calculation -- if $\epsilon_B/\epsilon_{\rm CR} = \beta$ = const is
somehow known (independent information about CRs electrons can come from X-ray data (Inverse Compton effect), or about
CRs from gamma rays (bremsstrahlung or pion decay)), system can be solved! It means that the magnetic
field energy density can be any constant fraction of the cosmic
ray energy density and the "equipartition" procedure will give
appropriate formulae for the estimation of the amount of the total
energy in a source and magnetic field strength, namely
\begin{equation}
B' = \Big( \frac{4\beta}{ \gamma+1}\Big)^{2/(\gamma +5)}B,
\end{equation}
where $B'$ is recalculated field for $\beta$=const, while $B$ is
the field corresponding to the minimum of energy. Total energy
calculated in this way is always higher than the minimal energy
obtained from the equipartition i.e. the minimum energy
calculation, but magnetic field can be either larger or
smaller.

Given the above, the equipartition calculation is not a precise method
for the determination of the magnetic field strength, but
we can surely estimate its order of magnitude (Duric 1990). The
main question is whether there is a physical relation between
$K_{\rm e}$ and $B$?  From Bell's (1978b) theory, $K_{\rm e}$
depends on the CR energy density $\epsilon_{\rm CR}$, injection
energy $E_{\rm inj}$ and the energy spectral index of cosmic ray particles
$\gamma$. Thus, implicitly, it must depend on shock velocity which
itself depends on time $t$ or radius $R=R(t)$ of an SNR. If there
is evolution of the magnetic field $B=B(t)$, $K_{\rm e}$ and $B$
must be related. Additionally, in the advanced model of DSA a
significant fraction of shock energy is transferred to CRs so the
cosmic ray pressure has to be included in equations (Drury 1983).
From this, the so-called non-linear DSA theory, the strong
magnetic field amplification (approximately two order of
magnitudes) is expected especially in the early free expansion
phase of SNR evolution, when the very strong shock waves exist
(Bell 2004). The non-linear effects thus make efficient cosmic ray
acceleration and, at the same time, the significant amplification
of the magnetic field strength. This increasing trend for the both
energy constituents of the synchrotron emission, again leads to
some form of a non-strict equipartition.

The derivation procedure presented in Subsection 2.1, where
integration limits for momenta are from 0 to $\infty$, leads to
equation (12). Using this equation and equation (3), the
calculated values of the magnetic field strength are slightly
overestimated (a few percent or more, depending on $E_{\mathrm{inj}}$). On the other hand, we neglect all kind of energy
losses in this paper. The main processes responsible for the
energy losses of the relativistic electrons are the synchrotron
radiation and the inverse Compton scattering. These energy losses
become significant for electrons especially at the very high
energies (radiation power for both processes depends on the square
of the electron kinetic energy). The energy losses of electrons
result in underestimation of the equipartition magnetic field
strength. Thus, in our "simple approach" (Subsection 2.1), the
effects of extending integration limits and energy losses work in
the opposite directions and may roughly cancel each other. If
integration limits are from $p_{\rm inj}$ to $\infty$ (Subsection
2.2), the equipartition calculation is derived correctly (without
assumption about the low shock velocity), but the problem of the
energy losses remains and the equipartition estimates fail for the
electrons at the highest energies (BK05). This discussion is
concentrated only on the energy losses of cosmic ray electrons.
The energy pool of cosmic rays is mainly filled with protons and
heavier particles which do not lose energy heavily by synchrotron
radiation and by inverse Compton scattering. Following Bell's
(1978b) theory, the energy ratio between cosmic ray protons and
electrons, for the energy spectral index $\gamma=2$, is approximately 40.
If we take $\gamma=2.5$, which is assumed for the obtained curves
presented in Figure 2, this ratio is $\approx7$. Due to this, the
total cosmic ray energy losses, in the first approximation, can be
neglected, especially for the objects with harder spectra (SNRs),
where the energy indices are lower\footnote{The average energy
index for SNRs is $\gamma\approx2$ (the radio spectral index
$\alpha\approx0.5$).}. However, the injection theory has been
developed for protons and heavy particles, but not for electrons
which may or may not follow the protons.
Hence Bell's formula may give only lower limits for the proton to electron
ratio at high energies and hence for the field strength.

In Table 1, we present values of the magnetic field strength and
the minimal energy for the sample of 30 Galactic SNRs for which
all data\footnote{Including the distances to SNRs independent of
the $\Sigma-D$ relation (see Uro\v{s}evi\'{c} et al. 2010 and
references therein), and spectral indices $0.5<\alpha<1$.}
necessary for the calculation can be found in the literature.
The calculated magnetic field strengths are close to those calculated by using revised equipartition
(BK05) and higher than those calculated by using classical equipartition (P70) for all 30 SNRs (see Figures 3 and 4).
For P70 calculation we used $\mathcal{K} = ({m_p}/{m_e})^{(3-\gamma)/2}$, $f=0.25$ and frequency interval $10^7\ \mathrm{Hz} < \nu < 10^{11}\ \mathrm{Hz}$.
For BK05 calculation, in order to convert from specific intensity to flux density we used $\frac{I_\nu}{l} = \frac{L_\nu}{4\pi V}$ ($L_\nu = 4\pi d^2 S_\nu$),\footnote{At the end of p.415 of their paper, BK05 suggested replacing $\frac{I_\nu}{l}$ with
$\frac{L_\nu}{V}$ which is incorrect, $4\pi$ is missing in denominator of the latter expression.}  $\mathbf{K_0} = ({m_p}/{m_e})^{\alpha}$ and $f=0.25$.
For five younger
Galactic SNRs, for which the forward shock velocities are known,
we use general equation for $\kappa$ (24). Differences between
calculated values, obtained by using general and simple
approaches, are generally not so high. If we define the fractional error
\begin{equation}
\varphi = \frac{|B - B_{v_s =0} |}{B_{v_s =0}},
\end{equation}
for the five SNRs with estimated shock velocities $\bar{\varphi}$ = 11\%.
For the youngest Galactic
SNR G1.9+0.3, the fractional error is the largest, $\varphi_\mathrm{max}$ = 30\%
(see Figure 4). Further inspection of Table 1, and Figures 3 and 4
leads to the conclusion that variation in abundances of CR species
does not significantly alter the final equipartition results.

\section{Conclusions}

In this paper we derived modified equipartition i.e.
minimum-energy formula for estimating magnetic fields in supernova
remnants. Our approach is similar to BK05 in a
sense that we do not integrate over frequencies as P70, however,
 \begin{description}
 \item [(i)] we assume power-law spectra $n(p) \propto
p^{-\gamma}$ and integrate over momentum to obtain energy
densities of particles,
 \item [(ii)] we take into account different ion species
 and not just equal number of protons and electrons at injection
 (e.g. for H to He ratio 10:1 there is more energy in $\alpha$-particles then in
 electrons),
\item [(iii)] we use flux density at a given frequency and also assume
isotropic distribution of the pitch angles for the remnant as a
whole,
\item[(iv)] by incorporating the dependence $\epsilon = \epsilon
(E_{\mathrm{inj}})$ we made the formula applicable to the younger
remnants as well.
\item[(v)] we calculate the magnetic field strengths for the sample of 30 Galactic SNRs and obtain values which are close to those calculated by using revised equipartition (BK05) and higher than those calculated by using classical equipartition (P70).
 \end{description}

\acknowledgments

We would like to thank anonymous referee for very useful comments and suggestions. 
During the work on this paper the authors were financially
supported by the Ministry of Education and Science of the Republic
of Serbia through the projects:  176004 'Stellar physics', 176005
'Emission nebulae: structure and evolution' and 176021 'Visible
and invisible matter in nearby galaxies: theory and observations'.

\clearpage

\begin{figure}
\epsscale{1.0} \plotone{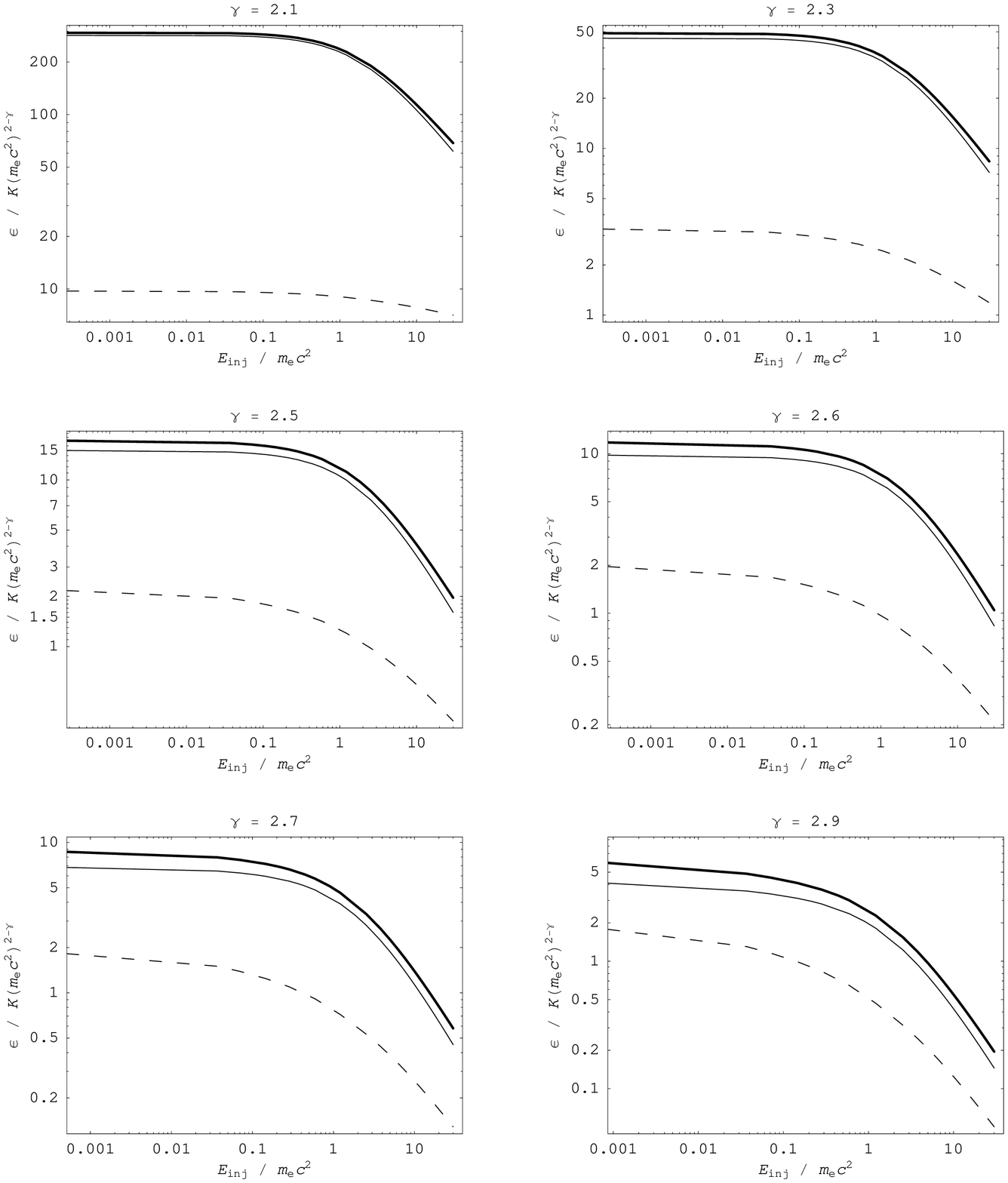} \caption{CR energy density of
ions (H:He = 10:1, solid line), electrons (dashed line) and total
(thick solid line) as a function of injection energy, in our
approximation.\label{fig1}}
\end{figure}

\clearpage

\begin{figure}
\epsscale{.75} \plotone{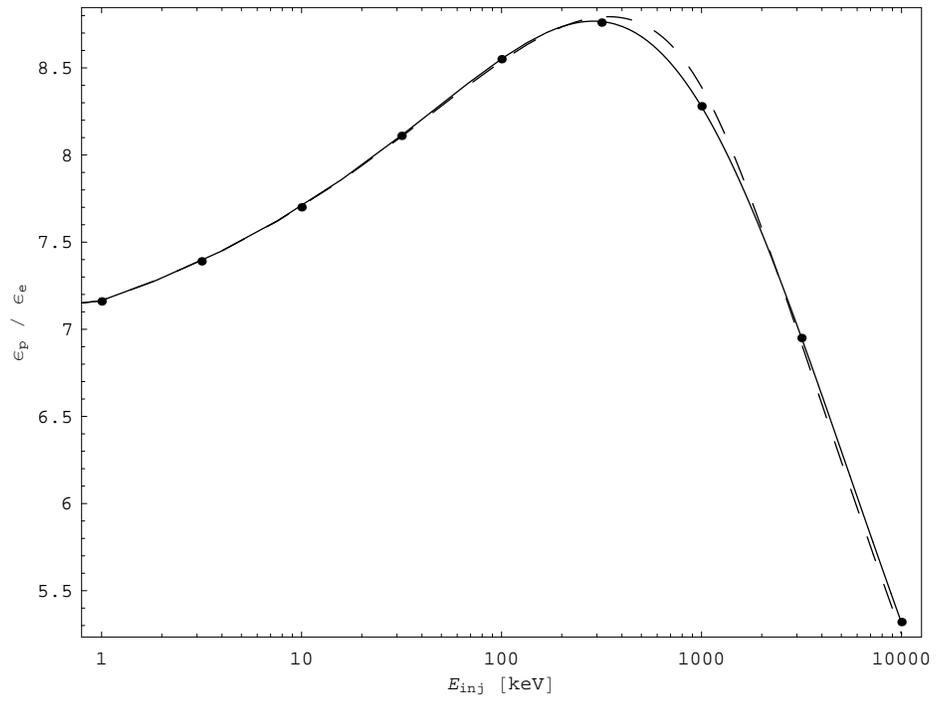} \caption{Proton to electron
energy density ratio as a function of injection energy in our
approximation (dashed line) and exact ratio (solid line) for $\gamma = 2.5$. Data points are
from Bell (1978b). \label{fig2}}
\end{figure}

\clearpage

\begin{figure}
\epsscale{.90} \plotone{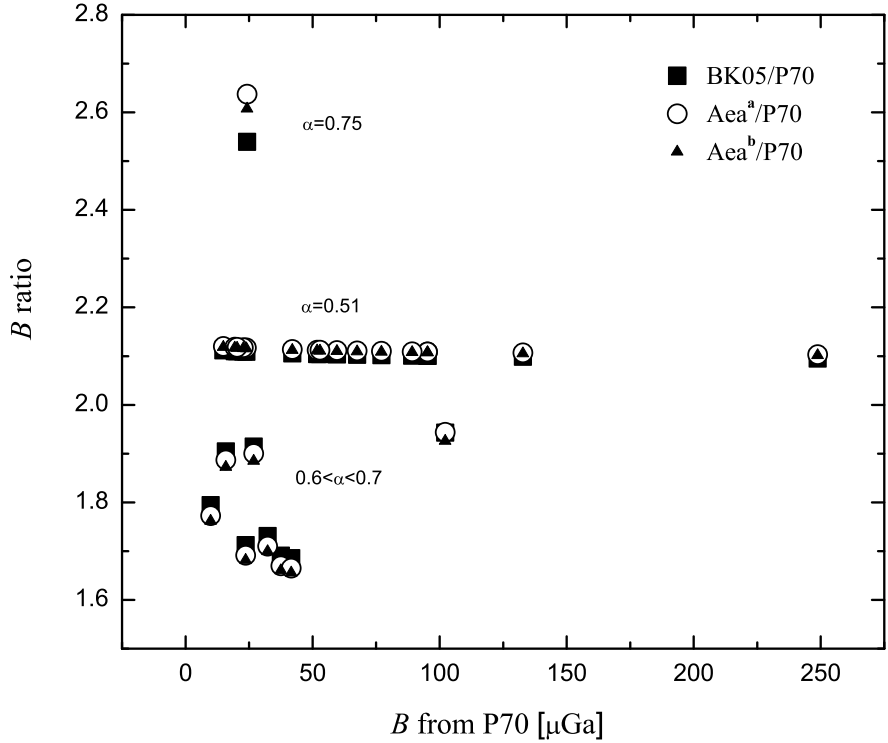} \caption{Comparison between different calculations for the minimum-energy
magnetic field strength ($B_{\rm{}}$). "$B_{\rm{}} \hspace{1mm} \rm{\small{ratio}}$" represents ratio between BK05 or this paper calculations,
and classical equipartition results (P70). Used abbreviations: Aea$^{\rm{a}}$ - this paper (Arbutina et al.), simple approach for p$^{\small{+}}$:e$^{\small{-}}$=1:1;
Aea$^{\rm{b}}$ - this paper, simple approach for H:He=10:1. Data are from Table 1 (25 SNRs). \label{fig3}}
\end{figure}

\clearpage

\begin{figure}
\epsscale{.90} \plotone{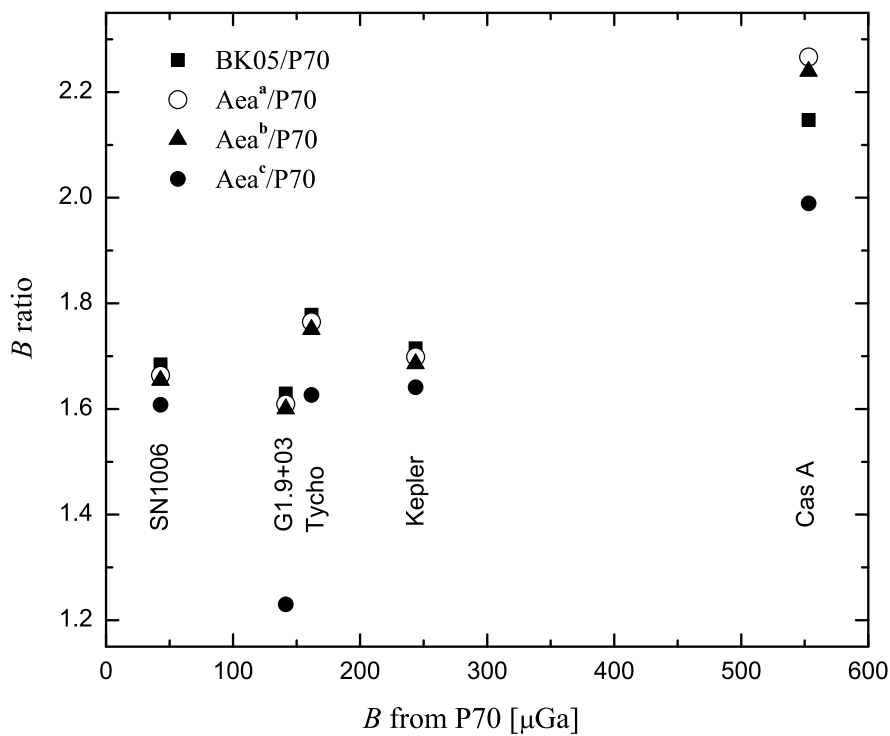} \caption{Comparison between different calculations for the minimum-energy
magnetic field strength ($B_{\rm{}}$) for 5 young SNRs with available forward shock velocities. "$B_{\rm{}} \hspace{1mm} \rm{\small{ratio}}$" represents ratio between BK05 or this paper calculations,
and classical equipartition results (P70). Used abbreviations: Aea$^{\rm{a}}$ - this paper (Arbutina et al.), simple approach for p$^{\small{+}}$:e$^{\small{-}}$=1:1;
Aea$^{\rm{b}}$ - this paper, simple approach for H:He=10:1; Aea$^{\rm{c}}$ - this paper, general approach for H:He=10:1.
Data are from Table 1 (5 SNRs). \label{fig4}}
\end{figure}

\clearpage

\addtolength{\hoffset}{0pt}
\begin{deluxetable}{@{\extracolsep{0mm}}llcccccccccc@{}}

\tablecolumns{12}
\tablewidth{0pc}
\rotate
\tabletypesize{\scriptsize}
\tablecaption{Calculated magnetic field strengths and total energies for sample of 30 Galactic SNRs}
\vskip -2mm
\tablehead{
\colhead{} & \colhead{} & \multicolumn{2}{c}{Pacholczyk(1970)}  & \multicolumn{2}{c}{Beck \& Krause (2005)}
& \multicolumn{2}{c}{This paper$^a$}  & \multicolumn{2}{c}{This paper$^b$} & \multicolumn{2}{c}{This paper$^c$} \\
\cline{3-4} \cline{5-6}  \cline{7-8}  \cline{9-10}  \cline{11-12} \\
\colhead{Name$^{d}$} & \colhead{Other names} & \colhead{$B$} & \colhead{$E_{\rm{min}}$}
& \colhead{$B$} & \colhead{$E_{\rm{min}}$}
& \colhead{$B$} & \colhead{$E_{\rm{min}}$}
& \colhead{$B$} & \colhead{$E_{\rm{min}}$}
& \colhead{$B$} & \colhead{$E_{\rm{min}}$}
}

\startdata

G4.5+6.8$^{e}$&  Kepler, SN1604, 3C358  & 2.44E-04	& 8.38E+47 & 4.18E-04  & 2.34E+48 & 4.14E-04  & 2.30E+48 &  4.11E-04 &   2.26E+48  & 4.00E-04 & 2.14E+48 \\
G21.8–0.6    &  Kes 69   &  7.71E-05 & 1.10E+50 & 1.62E-04	& 4.82E+50  &   1.63E-04   & 4.86E+50   & 1.63E-04 &   4.86E+50  &     -    &  -     \\
G23.3–0.3    &  W41      & 6.75E-05	& 5.86E+49 & 1.42E-04 &	2.57E+50 & 1.43E-04   & 2.59E+50   &  1.42E-04 &   2.59E+50  &     -      &   -   \\
G27.4+0.0    &  4C–04.71 & 1.02E-04 & 3.75E+48	& 1.99E-04	& 1.32E+49 &   1.99E-04   & 1.33E+49   &  1.97E-04 &   1.30E+49  &     -    &   -   \\
G33.6+0.1    &  Kes 79, 4C00.70, HC13   &  9.52E-05 &	3.79E+49 & 2.00E-04 & 1.66E+50  & 2.01E-04   & 1.68E+50   &  2.01E-04 &   1.67E+50 &  - &  - \\
G46.8–0.3    &  HC30 & 5.96E-05	& 4.88E+49	& 1.25E-04	& 2.14E+50 &   1.26E-04   & 2.16E+50   &  1.26E-04 &   2.16E+50  &      -      &     - \\ G53.6–2.2    & 3C400.2, NRAO 611   &  2.42E-05 &	3.18E+48 &	6.14E-05 &	1.88E+49   &   6.38E-05   & 2.02E+49   &  6.30E-05 &   1.98E+49  &  -  &  - \\
G65.1+0.6    &    -      &  9.90E-06	& 1.90E+50 &	1.78E-05	& 5.87E+50   &   1.76E-05   & 5.73E+50   &  1.74E-05 &   5.66E+50  & - &  -  \\
G93.7–0.2    &  CTB 104A, DA 551        &  2.68E-05 &	1.09E+49 &	5.13E-05	& 3.80E+49 &   5.09E-05   & 3.74E+49   &  5.05E-05 & 3.68E+49  &  - & - \\
G96.0+2.0    &   - &  1.49E-05 &	2.20E+48 &	3.15E-05 &	9.74E+48  &   3.16E-05   & 9.82E+48   &  3.16E-05 &   9.81E+48  &     -     &       -    \\
G108.2–0.6   &  -    &  1.94E-05	& 2.52E+49 &	4.09E-05 &	1.12E+50 &   4.11E-05   & 1.13E+50   &  4.11E-05 &   1.12E+50  &      -    &    -      \\
G109.1–1.0   &  CTB 109  &  5.18E-05 &	1.40E+49 &	1.09E-04 &	6.16E+49   &   1.09E-04   & 6.21E+49   &  1.09E-04 &   6.20E+49  &    -  &   -       \\
G111.7–2.1$^{f}$   &  Cassiopeia A, 3C461     &  5.53E-04 &	1.32E+49 &	1.19E-03 &	5.56E+49 &   1.25E-03   & 6.19E+49   &  1.24E-03 &   6.05E+49  & 1.10E-03   &   4.76E+49   \\
G114.3+0.3   &    -  &  2.40E-05 & 6.05E+47 &	5.05E-05 & 2.67E+48  &   5.07E-05   & 2.69E+48   &  5.07E-05 &   2.69E+48  &       -     &     -  \\
G116.5+1.1   & - &  2.27E-05 & 6.21E+48	& 4.80E-05 & 2.75E+49  &   4.82E-05   & 2.77E+49   &  4.81E-05 &   2.76E+49  &     -       &    -     \\
G116.9+0.2   &  CTB 1   &  3.23E-05 &	1.48E+48	& 5.60E-05 & 4.26E+48  &   5.53E-05   & 4.16E+48   &  5.49E-05 &   4.10E+48  &     -       &     -  \\
G120.1+1.4$^{g}$   &  Tycho, 3C10, SN1572     &  1.62E-04 & 1.63E+48 &	2.88E-04 &	4.88E+48   &   2.85E-04   & 4.80E+48   &  2.83E-04 &   4.73E+48  & 2.63E-04  &   4.09E+48   \\
G132.7+1.3   &  HB3  &  2.36E-05 & 2.69E+49 &	4.05E-05 &	7.58E+49   &   4.00E-05   & 7.39E+49   &  3.98E-05 & 7.31E+49  &  -  & -       \\
G160.9+2.6   &  HB9    &  1.58E-05 &	3.08E+50 &	3.02E-05 &	1.06E+51   &   2.99E-05   & 1.04E+51   &  2.97E-05 &   1.03E+51  &  -   &        -      \\
G205.5+0.5   &  Monoceros Nebula  & 2.03E-05 &	6.65E+49 &	4.27E-05 &	2.94E+50 &   4.29E-05   & 2.97E+50   &  4.29E-05 &   2.96E+50  &   -   & - \\
G260.4–3.4   &  Puppis A, MSH 08–44 &  5.29E-05 &	4.31E+49 &	1.11E-04 &	1.90E+50   &   1.12E-04   & 1.91E+50   &  1.12E-04 &   1.91E+50  &    -  & - \\
G292.2–0.5   &   - &  4.20E-05 &	4.78E+49 &	8.84E-05 &	2.11E+50   &   8.87E-05   & 2.12E+50   &  8.87E-05 &   2.12E+50  &     -      &        -    \\
G296.8–0.3   &  1156–62 & 3.75E-05 &	6.50E+49 &	6.34E-05 &	1.41E+50   &   6.26E-05   & 1.38E+50   &  6.22E-05 &   1.36E+50  &     -     &       -  \\
G304.6+0.1   &  Kes 17             &  9.52E-05 &	3.73E+49 &	2.00E-04 &	1.64E+50   &   2.01E-04   & 1.65E+50   &  2.01E-04 &   1.65E+50  & -  &  -  \\
G315.4–2.3   &  RCW 86, MSH 14–63   & 4.16E-05 &	1.37E+49 &	7.01E-05 &	3.75E+49  &   6.92E-05   & 3.66E+49   &  6.88E-05 &   3.62E+49  &  - &  - \\
G327.6+14.6$^{h}$  &  SN1006, PKS 1459–41     &  4.28E-05 &	4.65E+48  & 7.22E-05 & 1.27E+49  &   7.13E-05   & 1.24E+49   &  7.09E-05 &
1.22E+49  & 6.89E-05   &   1.16E+49   \\
G332.4–0.4   &  RCW 103    &  1.33E-04 &	4.63E+48 &	2.79E-04 &	2.03E+49   &   2.80E-04   & 2.04E+49   &  2.80E-04 &   2.04E+49  &   -    &  -      \\
G337.8–0.1   &  Kes 41     &  8.92E-05 & 6.80E+49 &	1.87E-04 &	2.99E+50 &   1.88E-04   & 3.01E+50   &  1.88E-04 &   3.00E+50  &     -   &   -    \\
G349.7+0.2   &        -   &  2.49E-04 & 6.49E+49 &	5.21E-04 &	2.83E+50   &   5.23E-04   & 2.85E+50   &  5.23E-04 &   2.85E+50  &      -      &  -   \\
G1.9+03$^{i}$ &   -  & 1.41E-04 &	3.65E+47 &	2.30E-04 &	9.33E+47  &   2.28E-04   & 9.11E+47   &  2.26E-04 &   9.00E+47  & 1.74E-04   &  5.31E+47   \\

\enddata
\vskip 0mm
\tablecomments{All units are in CGS system. $B$ is magnetic field strength calculated for minimum-energy assumption.}
\tablenotetext{a}{Simple approach for p$^{\small{+}}$:e$^{\small{-}}$=1:1.}
\tablenotetext{b}{Simple approach for H:He=10:1.}
\tablenotetext{c}{General approach for H:He=10:1, for young SNRs with available forward shock velocities ($\upsilon_{\rm{s}}$).}
\tablenotetext{d}{According to Green's (2009) catalogue from which data for SNRs, except shock velocities, has been taken. }
\tablenotetext{e}{$\upsilon_{\rm{s}} = 1660 \rm{km/s}$ (Sankrit et al. 2005).}
\tablenotetext{f}{$\upsilon_{\rm{s}} = 4900 \rm{km/s}$ (Patnaude et al. 2009).}
\tablenotetext{g}{$\upsilon_{\rm{s}} = 4700 \rm{km/s}$ (Hayato et al. 2011).}
\tablenotetext{h}{$\upsilon_{\rm{s}} = 2890 \rm{km/s}$ (Ghavamian et al. 2002).}
\tablenotetext{i}{$\upsilon_{\rm{s}} = 14000 \rm{km/s}$ (Carlton et al. 2011).}

%\tablenotetext{\ }{$^a$Simple approach, p$^{+}:e^{-}=1:1$, $^b$Simple approach, H:He=10:1, $^c$General approach, H:He=10:1, for young SNRs with available\
% shock %velocities, $^{d}$ according to Green's (2009) catalogue from which data for SNRs has been taken, except for shock velocities which has been taken from $^{e}$
%Sankrit et al. (2005), $^{f}$ Patnaude et al. (2009), $^{g}$ Hayato et al. (2011), $^{h}$ Ghavamian et al. (2002) and $^{i}$ Carlton et al. (2011)}

\end{deluxetable}

\end{document}